\documentclass[conference]{IEEEtran}
\IEEEoverridecommandlockouts
\usepackage{cite}
\usepackage{amsmath,amssymb,amsfonts}
\usepackage{algorithmic}
\usepackage{graphicx}
\usepackage{textcomp}
\usepackage{xcolor}
\usepackage{booktabs}
\usepackage{multirow}
\usepackage{hyperref}
\usepackage{array}
\usepackage{tabularx}
\def\BibTeX{{\rm B\kern-.05em{\sc i\kern-.025em b}\kern-.08em
    T\kern-.1667em\lower.7ex\hbox{E}\kern-.125emX}}
\begin{document}

\title{Context-Aware Mapping of 2D Drawing Annotations to 3D CAD Features Using LLM-Assisted Reasoning for Manufacturing Automation\\
}

\author{\IEEEauthorblockN{Muhammad Tayyab Khan\textsuperscript{a, c*}, Lequn Chen\textsuperscript{b}, Wenhe Feng\textsuperscript{a}, Seung Ki Moon\textsuperscript{c}}
\IEEEauthorblockA{\textsuperscript{a} Singapore Institute of Manufacturing Technology (SIMTech), A*STAR, Singapore 636732, Republic of Singapore}
\IEEEauthorblockA{\textsuperscript{b} Advanced Remanufacturing and Technology Centre (ARTC), A*STAR, Singapore 637143, Republic of Singapore}
\IEEEauthorblockA{\textsuperscript{c} School of Mechanical and Aerospace Engineering, Nanyang Technological University, Singapore 639798}
\IEEEauthorblockA{*Email: KHAN0022@e.ntu.edu.sg}
}

\maketitle

\begin{abstract}
Manufacturing automation in process planning, inspection planning, and digital-thread integration depends on a unified specification that binds the geometric features of a 3D CAD model to the geometric dimensioning and tolerancing (GD\&T) callouts, datum definitions, and surface requirements carried by the corresponding 2D engineering drawing. Although Model-Based Definition (MBD) allows such specifications to be embedded directly in 3D models, 2D drawings remain the primary carrier of manufacturing intent in automotive, aerospace, shipbuilding, and heavy-machinery industries. Correctly linking drawing annotations to the corresponding 3D features is difficult because of contextual ambiguity, repeated feature patterns, and the need for transparent and traceable decisions. This paper presents a deterministic-first, context-aware framework that maps 2D drawing entities to 3D CAD features to produce a unified manufacturing specification. Drawing callouts are first semantically enriched and then scored against candidate features using an interpretable metric that combines type compatibility, tolerance-aware dimensional agreement, and conservative context consistency, along with engineering-domain heuristics. When deterministic scoring cannot resolve an ambiguity, the system escalates to multimodal and constrained large-language-model reasoning, followed by a single human-in-the-loop (HITL) review step. Experiments on 20 real CAD--drawing pairs achieve a mean precision of 83.67\%, recall of 90.46\%, and F1 score of 86.29\%. An ablation study shows that each pipeline component contributes to overall accuracy, with the full system outperforming all reduced variants. By prioritizing deterministic rules, clear decision tracking, and retaining unresolved cases for human review, the framework provides a practical foundation for downstream manufacturing automation in real-world industrial environments.
\end{abstract}


\section{Introduction}

Modern manufacturing automation relies on the accurate and machine-actionable interpretation of design intent~\cite{ref1}. While 3D CAD models provide precise geometric descriptions and enable automatic feature recognition (AFR), they do not fully capture manufacturing requirements~\cite{ref2}. Critical information, such as geometric dimensioning and tolerancing (GD\&T) constraints, datum references, surface finish requirements, and thread specifications are typically specified in 2D drawings~\cite{ref3,ref4}. Downstream tasks, including process planning, inspection planning, tolerance analysis, and digital-thread integration, require these heterogeneous representations to be unified into a consistent manufacturing specification. Notably, despite advances in Model-Based Definition (MBD)~\cite{ref6}, 2D drawings remain the authoritative source of symbolic manufacturing semantics in many industrial workflows~\cite{ref5,ref7}.

Aligning 2D drawing intent with 3D geometry remains challenging because drawing annotations are inherently context dependent. A numeric value such as ``10'' may denote a hole diameter, a slot width, a fillet radius, or a depth depending on drafting conventions, leader placement, and view association. GD\&T symbols further complicate interpretation because visually similar annotations may impose fundamentally different constraints depending on symbol type, modifiers, and datum references. Misinterpretation at this stage can propagate errors into manufacturing and inspection planning, undermining automation reliability.

As detailed in Section~II, despite advances in drawing understanding, AFR, and LLM-based engineering tools, no integrated method exists for mapping drawing-level manufacturing semantics to recognized 3D features with transparent, auditable decision-making. Deterministic matching strategies that rely solely on numeric equality are unreliable under ambiguity, while purely data-driven approaches can be opaque and difficult to audit.

To address this gap, we propose a context-aware mapping framework that uses a VLM to enrich drawing callouts into structured semantic descriptors, followed by deterministic scoring and selection governed by explicit engineering rules. The approach emphasizes reproducibility, conservative spatial reasoning, and strict GD\&T disambiguation, making it suitable for deployment in industrial manufacturing automation pipelines. To the best of our knowledge, this is the first study to automatically and contextually map 2D drawing information to CAD features using AI-assisted reasoning while preserving deterministic, transparent, and auditable decision-making required for industrial manufacturing workflows.

The main contributions of this paper are as follows:
\begin{enumerate}
    \item We formally define the 2D--3D manufacturing intent mapping problem with explicit input and output representations, establishing a precise formulation for cross-modal correspondence.
    \item We propose a deterministic-first escalation framework that combines interpretable correspondence scoring with engineering-aware heuristics and controlled LLM-assisted reasoning, enabling transparent, auditable, and manufacturing-grade mapping decisions.
    \item We establish an evaluation protocol and present benchmark results on real industrial parts, including ablation experiments that quantify the contribution of each pipeline component.
\end{enumerate}

The remainder of this paper is organized as follows. Section~II reviews related work. Section~III formulates the 2D--3D manufacturing intent mapping problem. Section~IV presents the proposed deterministic-first mapping framework. Section~V describes the experimental setup and quantitative results. Section~VI discusses performance and limitations. Finally, Section~VII concludes the paper.

\section{Related Work}

Prior work on drawing understanding has focused on symbol detection, OCR, and syntactic parsing to recover structured dimensions and GD\&T frames~\cite{ref8,ref9}. Khallouli et al.~\cite{ref5} demonstrated transformer-based OCR with generative data augmentation for engineering document recognition. Khan et al.~\cite{ref2,ref3,ref4} proposed hybrid vision--language frameworks for extracting structured manufacturing knowledge from multi-view engineering drawings. While these methods successfully extract individual entities from drawings, they treat them as isolated annotations and do not establish correspondences between extracted drawing entities and specific 3D geometric features.

In parallel, AFR research extracts manufacturable features from CAD models using geometric reasoning or learning-based approaches. Zhang et al.~\cite{ref11} introduced FeatureNet for 3D convolutional feature recognition, Wu et al.~\cite{ref12} proposed AAGNet using graph neural networks, and Khan et al.~\cite{ref10} leveraged vision--language models (VLMs) for feature recognition. These methods operate exclusively on 3D geometry and do not incorporate manufacturing semantics from 2D drawings. MBD aims to embed manufacturing information directly within 3D models~\cite{ref6}, but adoption remains limited in practice~\cite{ref7}; many industries continue to rely on 2D drawings due to legacy workflows, certification requirements, and supply chain interoperability. The emerging application of LLMs and VLMs to engineering tasks has shown promise for information extraction, code generation, and design reasoning~\cite{ref13,ref14}. Eisert et al.~\cite{ref15} recently proposed a pipeline for converting 2D drafting annotations into 3D PMI using object detection, but their method operates at the annotation level and does not perform feature-level correspondence. No prior work has applied AI-assisted reasoning to establish correspondences between 2D drawing annotations and 3D CAD features. While each of these research areas addresses a component of the manufacturing interpretation pipeline, none provides an integrated, auditable method for linking drawing-level semantics to recognized 3D features with transparent decision-making.

\section{Problem Formulation}

We assume two pre-extracted structured sets as input, corresponding to CAD features and drawing entities.

\subsection{3D AFR Feature Set}

Let the set of extracted 3D CAD features be defined as
\begin{equation}
F_{3D} = \{f_i\}_{i=1}^{N} \;;\; f_i = (t_i, p_i, c_i, k_i)
\end{equation}
where
\begin{itemize}
    \item $t_i \in T$: feature type (e.g., hole, slot, pocket, fillet, etc.)
    \item $p_i \in \mathbb{R}^d$: geometric parameters (e.g., diameter, depth, radius, angle, etc.)
    \item $c_i \in [0, 1]$: AFR confidence score
    \item $k_i$: optional metadata (e.g., pattern ID, symmetry group, etc.)
\end{itemize}
Each feature may additionally include spatial information such as a centroid $c(f_i) = (x_i, y_i, z_i)$ or a bounding box.

\subsection{2D Drawing Entity Set}

Let the set of extracted 2D drawing entities be defined as
\begin{equation}
E_{2D} = \{e_j\}_{j=1}^{M} \;;\; e_j = (u_j, q_j, \tau_j, \gamma_j)
\end{equation}
where
\begin{itemize}
    \item $u_j \in U$: entity type (e.g., dimension, GD\&T, surface roughness symbol, notes, etc.)
    \item $q_j$: parsed semantic values (e.g., \{value: 10, unit: mm, type: diameter\}, etc.)
    \item $\tau_j$: raw OCR text
    \item $\gamma_j$: local context (e.g., bounding box, neighboring annotations, associated drawing views, etc.)
\end{itemize}

\subsection{Output Specification}

The output of the proposed system consists of
\begin{enumerate}
    \item \textbf{Mapping set} $M \subseteq F_{3D} \times E_{2D}$, where each pair is associated with a confidence score and a rationale.
    \item \textbf{Unified manufacturing specification}, represented as a structured document that binds drawing constraints to CAD features, explicitly lists unmapped entities, and records provenance information including mapping method, confidence, and any human edits.
\end{enumerate}

\section{Methodology}

This section presents the proposed deterministic-first, context-aware framework for mapping drawing entities to CAD features. The method integrates semantic enrichment, deterministic correspondence scoring, engineering-aware heuristics, and controlled escalation mechanisms, followed by a single post-fusion human-in-the-loop (HITL) validation step. An overview of the complete pipeline is shown in Fig.~1.

\begin{figure}[t]
\centering
\includegraphics[width=\columnwidth]{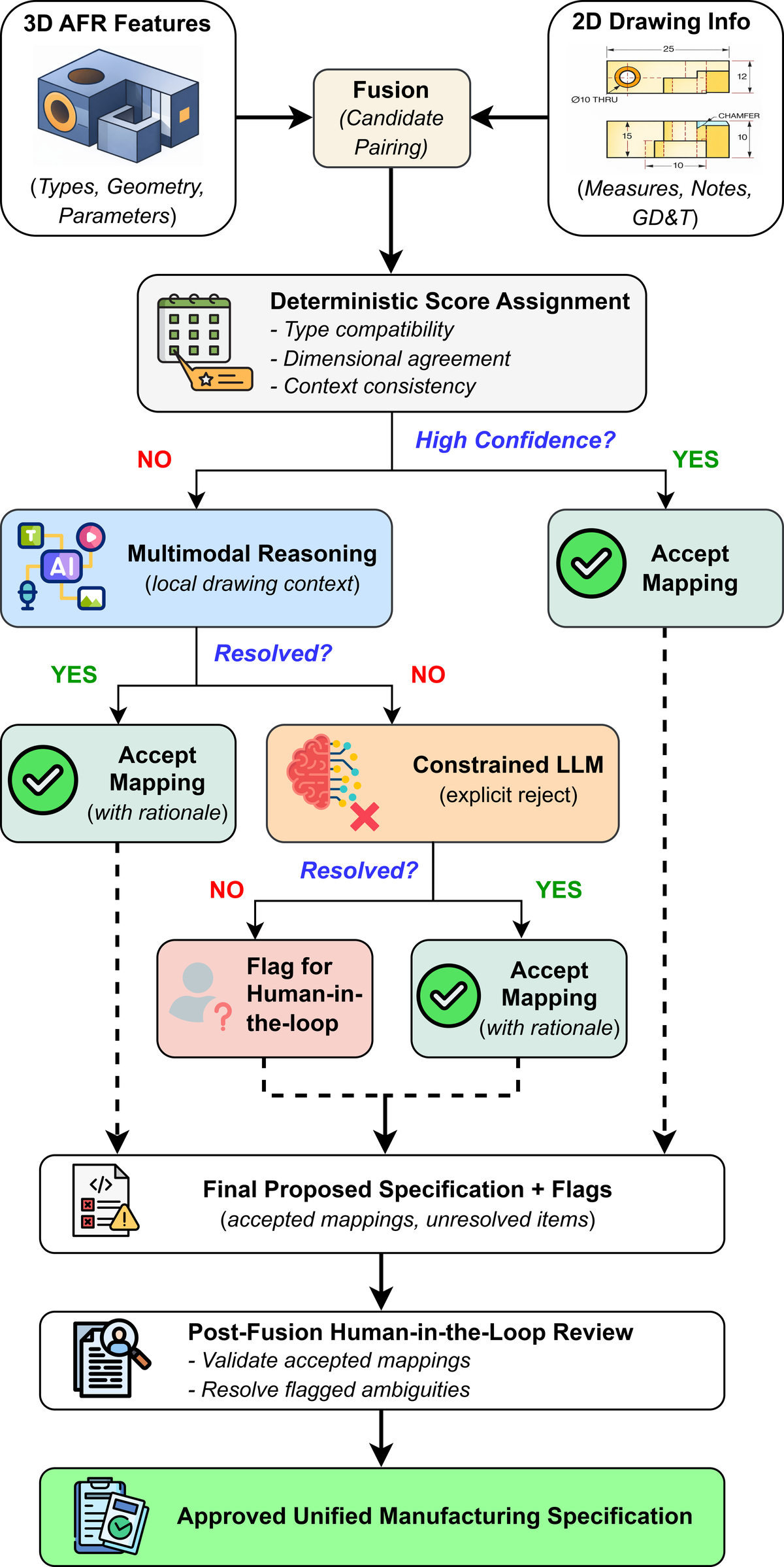}
\caption{Overview of the deterministic-first, context-aware fusion pipeline for mapping 2D drawing entities to 3D CAD features.}
\label{fig:pipeline}
\end{figure}

\subsection{Overview of the Context Mapping Pipeline}

As illustrated in Fig.~1, the pipeline takes as input pre-extracted CAD features and 2D drawing entities, which are fused to generate candidate feature--entity correspondences. The pipeline proceeds through the following stages:
\begin{enumerate}
    \item \textbf{Deterministic scoring and assignment} based on semantic compatibility, tolerance-aware dimensional agreement, and conservative context consistency.
    \item \textbf{Engineering-aware heuristic refinement} integrated within the deterministic scoring stage to address high-impact manufacturing cases.
    \item \textbf{Controlled escalation} to multimodal reasoning when deterministic confidence is insufficient, followed by generation of a proposed specification containing accepted mappings and unresolved entities.
    \item \textbf{A single post-fusion HITL review step} in which a human reviewer validates accepted mappings, resolves flagged ambiguities, and approves the final unified manufacturing specification.
\end{enumerate}

\subsection{Semantic Enrichment via VLM}

Before scoring, each raw drawing entity $e_j$ is semantically enriched by a VLM. The model receives a cropped image region around the annotation together with the OCR text $\tau_j$ and bounding-box coordinates $\gamma_j$, and returns a structured descriptor containing: (i)~a normalized entity type (e.g., ``diameter'', ``radius'', ``thread''), (ii)~parsed numeric values with units, (iii)~an inferred target feature category (e.g., ``hole'', ``fillet''), and (iv)~a confidence score $c_j^{VLM} \in [0,1]$ reflecting the model's certainty. This enrichment converts ambiguous OCR tokens such as ``10'' or ``M8'' into typed, semantically grounded descriptors that the downstream scoring functions can consume directly. Enrichment is performed once per entity before the scoring stage; the resulting descriptors and confidence scores are then used by the type compatibility (Eq.~4), dimensional agreement (Eq.~5), and context consistency (Eq.~6) components described below.

\subsection{Deterministic Correspondence Scoring}

For each pair $(f_i, e_j)$ consisting of a 3D feature and a 2D drawing entity, a composite correspondence score $S_{ij}$ is computed as
\begin{equation}
S_{ij} = \begin{cases} 0, & \text{if } S_{ij}^{type} = 0 \\[4pt] w_t S_{ij}^{type} + w_d S_{ij}^{dim} + w_c S_{ij}^{ctx} + h_{ij}, & \text{otherwise} \end{cases}
\end{equation}
subject to $w_t + w_d + w_c = 1$, where $h_{ij}$ denotes engineering heuristic adjustments (Section~IV-E). The resulting score is not normalized to $[0,1]$ and is used only for relative ranking within each feature's candidate set. The hard gate on $S_{ij}^{type} = 0$ ensures that type-incompatible pairs are rejected immediately, preventing numeric coincidences from producing false correspondences.

In this study, the weights are fixed as $w_t = 0.4$, $w_d = 0.4$, $w_c = 0.2$, and a candidate correspondence is retained only if $S_{ij} \geq \theta_{cand}$, where $\theta_{cand} = 0.3$. These values, together with $\rho = 0.9$ and $\varepsilon = 0.1$\,mm, were selected through preliminary experiments on a held-out set of five parts and remained fixed across all subsequent evaluations.

\subsubsection{Type Compatibility Modeling}
It determines whether a given 2D entity can meaningfully constrain a specific 3D feature and is defined as
\begin{equation}
S_{ij}^{type} = \begin{cases} 1.0, & \text{if } (t_i, u_j) \in \Omega_{exact} \\ 0.9, & \text{if } (t_i, u_j) \in \Omega_{semantic} \\ 0, & \text{otherwise} \end{cases}
\end{equation}
where $\Omega_{exact}$ denotes direct type matches (e.g., ``hole'' $\leftrightarrow$ ``hole'') and $\Omega_{semantic}$ denotes semantically equivalent groupings (e.g., \{hole, bore, drill\}, \{slot, pocket, groove\}, \{fillet, round, radius\}). This step eliminates invalid correspondences prior to numeric or contextual evaluation.

\subsubsection{Dimensional Agreement}
It evaluates consistency between numeric values extracted from the drawing and geometric parameters of the 3D feature. Matching is context-aware: the comparison is routed to the geometrically appropriate 3D property based on the semantic type of the 2D entity (e.g., a diameter dimension is compared only against the 3D diameter, not against width or depth). For a scalar attribute, the dimensional agreement score is defined as
\begin{equation}
S_{ij}^{dim} = \begin{cases} 1.0, & \text{if } |x_j^{2D} - x_i^{3D}| \leq \varepsilon \\ 0.7, & \text{if } |x_j^{2D} - x_i^{3D}| \leq 2\varepsilon \\ 0, & \text{otherwise} \end{cases}
\end{equation}
where $\varepsilon = 0.1$\,mm is the base dimensional tolerance. For radius entities, the score also checks $|x_j^{2D} - d_i^{3D}/2| \leq \varepsilon$, where $d_i^{3D}$ is the 3D diameter. An additional gate applies: if a numeric 2D dimension exists but $S_{ij}^{dim} = 0$, the composite score is multiplied by 0.3 after all additive components (including $h_{ij}$) have been computed, i.e., $S_{ij} \leftarrow 0.3 \cdot S_{ij}$. This strongly suppresses dimensionally mismatched correspondences.

\subsubsection{Context Consistency}
It incorporates spatial and view-related cues extracted from the drawing context, operationalized through the VLM confidence score:
\begin{equation}
S_{ij}^{ctx} = \begin{cases} 0.5, & \text{if spatial cues unavailable} \\ c_j^{VLM}, & \text{otherwise} \end{cases}
\end{equation}
where $c_j^{VLM} \in [0,1]$ is the VLM-reported confidence for the semantic enrichment of entity $e_j$. The neutral value of 0.5 reflects the absence of reliable spatial evidence. By design, $S_{ij}^{ctx}$ is conservative and does not override semantic compatibility or dimensional agreement.

\subsection{Assignment and Near-Tie Filtering}

After candidate filtering using $S_{ij} \geq \theta_{cand}$, assignment is performed per 3D feature using a greedy near-tie strategy. For each feature $f_i$, all candidate entities are ranked by score, and all entities whose scores lie within a fixed ratio $\rho$ of the best score are retained:
\begin{equation}
A_i = \{e_j \mid S_{ij} \geq \rho \cdot \max_k S_{ik}\} \;;\; \rho = 0.9
\end{equation}
This near-tie rule supports one-to-many correspondences that commonly arise when the same feature is dimensioned across multiple drawing views or when multiple annotations constrain a single feature (e.g., a counterbore with separate diameter and depth callouts). A single 2D entity may be assigned to multiple 3D features when the same dimension constrains multiple instances.

\subsection{Engineering-Aware Heuristic Refinement}

Engineering-aware heuristics are integrated within the deterministic scoring to restrict candidate sets, refine tolerance interpretation, and enforce domain-specific rules. The heuristic adjustment $h_{ij}$ in Eq.~(3) comprises:
\begin{itemize}
    \item \textbf{Diameter symbol preference:} For hole features, dimensions bearing the diameter symbol (\O) receive a bonus ($h_{ij} = +0.1$), while dimensions labeled as ``diameter'' by the VLM but lacking the \O{} symbol receive a multiplicative penalty ($S_{ij} \leftarrow 0.7 \cdot S_{ij}$).
    \item \textbf{Thread callouts:} For thread designations of the form $M(d)$, candidate features are restricted to cylindrical features. Agreement is scored using the same stepped tolerance matching with the nominal thread diameter compared against the 3D hole diameter.
    \item \textbf{Pattern multiplicity:} Drawing indications such as $n\text{X}$ permit mapping to a group of $n$ geometrically similar features.
    \item \textbf{GD\&T attachment priors:} Position and profile constraints are preferentially attached to holes and pockets, runout-related constraints to cylindrical features, and datum references to planar or cylindrical primitives.
\end{itemize}

\subsection{Escalation and Post-Fusion HITL Validation}

When deterministic confidence is insufficient, the pipeline escalates to multimodal reasoning using a localized drawing image around entity $e_j$ and the candidate feature list to resolve spatial ambiguity. If ambiguity persists, a constrained language model is invoked with a fixed output schema consisting of a decision, target feature identifier, confidence score, and evidence-based rationale.

\subsubsection{Model and Prompt Details}

The multimodal reasoning stage employs GPT-4o (OpenAI, version \textit{gpt-4o-2024-08-06}) as the VLM for both the multimodal escalation and constrained LLM reasoning stages. In the multimodal stage, the model receives a cropped drawing region centered on the ambiguous entity along with a structured JSON description of candidate 3D features, and is prompted to select the most likely correspondence or explicitly reject all candidates. The constrained LLM stage uses a structured output schema enforcing a fixed JSON format: \texttt{\{decision, target\_feature\_id, confidence, rationale\}}. Rejection is explicitly permitted and encouraged when evidence is insufficient.

\subsubsection{Escalation Statistics}
Across the 20-part evaluation dataset (101 total mappings), the pipeline resolved correspondences as follows: 4.0\% were resolved by deterministic scoring alone (single high-confidence candidate), 84.1\% through the combined deterministic-VLM pathway (deterministic scoring produced the candidate set, and the VLM selected among near-tie candidates), and 11.9\% required constrained LLM reasoning. Of the LLM-escalated cases, one mapping (1.0\% of total) was explicitly rejected due to insufficient evidence. These statistics show that deterministic scoring always produces the initial candidate ranking, but VLM-assisted selection is needed for the majority of cases because drawing annotations are inherently ambiguous.

\medskip
All unresolved or low-confidence cases are flagged for post-fusion HITL review. Provenance records automated decisions, human edits, and review effort, ensuring traceability and manufacturing-grade correctness.

\subsection{Unified Manufacturing Specification}

The output is a structured unified manufacturing specification that binds 2D drawing constraints to 3D CAD features with explicit mapping metadata and provenance as shown in Fig.~2. Unresolved entities are retained for review, enabling measurable performance evaluation and guaranteeing final correctness for downstream manufacturing and inspection tasks.

\begin{figure}[t]
\centering
\includegraphics[width=\columnwidth]{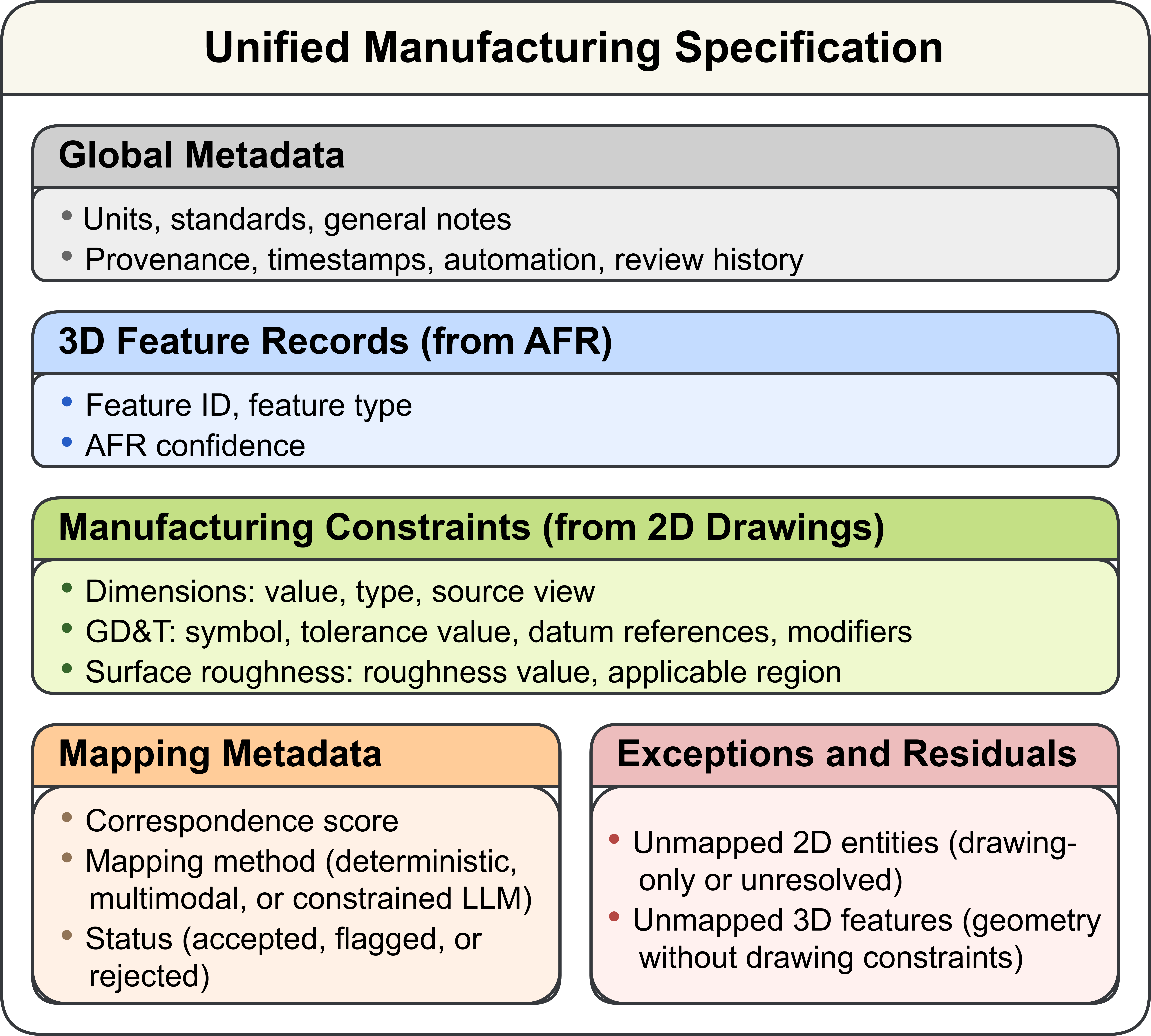}
\caption{Structure of the unified manufacturing specification binding 2D drawing constraints to 3D CAD features with mapping metadata and provenance.}
\label{fig:unified_spec}
\end{figure}

\section{Evaluation}

\subsection{Dataset and Ground Truth}

Evaluation was conducted on a dataset of 20 mechanical parts, each consisting of a CAD model and a corresponding 2D drawing. AFR for 3D geometry and entity extraction for drawings were performed upstream. Ground truth mappings between extracted 2D entities and 3D features were created manually by domain experts and used to evaluate the proposed method. The dataset spans parts with varying geometric complexity, feature repetition, and annotation density, reflecting realistic industrial drawing practices. All evaluations were performed using a single, consistent pipeline configuration, without part-specific assumptions or manual intervention during automated mapping.

Table~\ref{tab:dataset_stats} summarizes the key characteristics of the evaluation dataset in terms of geometric and annotation complexity per part.

\begin{table}[t]
\centering
\caption{Dataset Characterization: Complexity Statistics Across 20 CAD--Drawing Pairs.}
\label{tab:dataset_stats}
\begin{tabular}{lcccc}
\toprule
\textbf{Characteristic} & \textbf{Min} & \textbf{Max} & \textbf{Mean} \\
\midrule
3D features per part        & 2   & 9   & 5.1  \\
2D entities per part        & 2   & 13  & 5.6  \\
Repeated/patterned features & 0   & 6   & 2.0  \\
GD\&T annotations per part  & 0   & 1   & 0.05 \\
\bottomrule
\end{tabular}
\end{table}

Approximately 40\% of all 3D features (40 out of 101) are repeated or patterned instances, with through holes being the dominant type (37 out of 101). The dataset is predominantly dimension-centric; GD\&T annotations are essentially absent (only 1 out of 101 mappings), meaning the framework's GD\&T heuristics are not fully exercised in this evaluation. Ground truth annotations were created by two domain experts with disagreements adjudicated by a third reviewer.

\subsection{Evaluation Metrics}

Mapping performance is evaluated at the 2D link level. For each 3D feature, the set of predicted 2D entity associations is compared against the ground truth set via set intersection. Let $M$ denote the predicted mapping set and $M^*$ denote the ground truth mapping set. Precision, recall, and F1 score are defined as
\begin{equation}
P = \frac{|M \cap M^*|}{|M|},\; R = \frac{|M \cap M^*|}{|M^*|},\; F_1 = \frac{2PR}{P + R}
\end{equation}

Per-part metrics are computed and then macro-averaged across all 20 parts. In addition, exact match rate (fraction of 3D features with perfect 2D set correspondence) and partial match rate (fraction with at least one correct 2D entity) are reported.

\subsection{Case Study and Quantitative Results}

Fig.~3 presents a representative qualitative result. Fig.~3(a) shows the CAD model with recognized features (F1--F4), Fig.~3(b) shows the corresponding 2D drawing with extracted entities (E1--E4), and Fig.~3(c) illustrates the unified mapping after scoring, heuristic refinement, and controlled escalation. Deterministic correspondence is achieved for unambiguous cases such as the cylindrical boss (F1$\leftrightarrow$E1) and the fillet radius (F4$\leftrightarrow$E4), where strong type compatibility and dimensional agreement are present. More ambiguous cases involving repeated hole patterns (F2$\leftrightarrow$E2) and profile--thickness relationships in gusset features (F3$\leftrightarrow$E3) are resolved through hybrid reasoning with optional HITL validation.

\begin{figure*}[t]
\centering
\includegraphics[width=\textwidth]{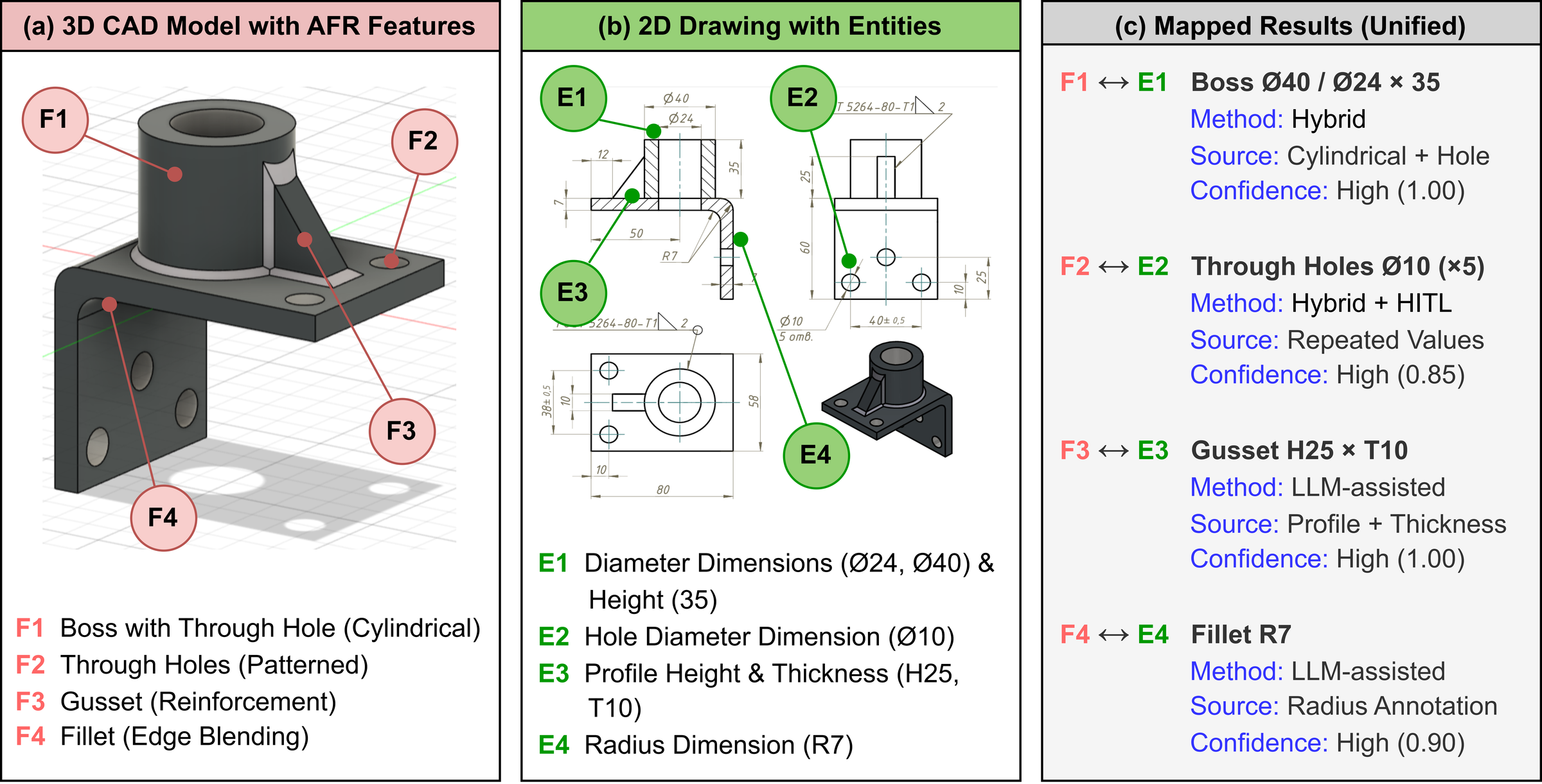}
\caption{Representative result of the proposed framework. (a)~3D CAD model with recognized features (F1--F4). (b)~Corresponding 2D engineering drawing with extracted entities (E1--E4). (c)~Unified mapping showing resolved correspondences with mapping method, source, and confidence.}
\label{fig:case_study}
\end{figure*}

Table~\ref{tab:main_results} summarizes the aggregated evaluation results across all 20 parts.

\begin{table}[t]
\centering
\caption{Aggregated Mapping Accuracy, Robustness, and Inference Time Statistics Across 20 CAD--Drawing Pairs.}
\label{tab:main_results}
\begin{tabular}{lcccc}
\toprule
\textbf{Metric} & \textbf{Mean} & \textbf{Std} & \textbf{Min} & \textbf{Max} \\
\midrule
Mapping Precision    & 0.8367 & 0.1698 & 0.50   & 1.0    \\
Mapping Recall       & 0.9046 & 0.1165 & 0.5714 & 1.0    \\
Mapping F1 score     & 0.8629 & 0.1367 & 0.5714 & 1.0    \\
Exact Match Rate     & 0.7919 & 0.1996 & 0.40   & 1.0    \\
Partial Match Rate   & 0.9025 & 0.1227 & 0.5714 & 1.0    \\
Inference Time (s)   & 54.94  & 20.65  & 30.05  & 104.63 \\
\bottomrule
\end{tabular}
\end{table}

\subsection{Ablation Study}

To quantify the contribution of each pipeline component, we conducted an ablation study by re-implementing the scoring pipeline in degraded configurations and evaluating each variant against the same ground truth. Table~\ref{tab:ablation} reports mean precision, recall, and F1 score for each variant across the 20-part dataset.

\begin{table}[t]
\centering
\caption{Ablation Study: Impact of Removing Individual Pipeline Components on Mapping Performance (Mean Across 20 Parts).}
\label{tab:ablation}
\begin{tabular}{lccc}
\toprule
\textbf{Variant} & \textbf{P} & \textbf{R} & \textbf{F1} \\
\midrule
Deterministic only          & 0.5088 & 0.3481 & 0.4057 \\
No engineering heuristics   & 0.4423 & 0.8867 & 0.5470 \\
No LLM escalation           & 0.7222 & 0.5417 & 0.6066 \\
No context scoring ($w_c{=}0$) & 0.6495 & \textbf{0.9093} & 0.7432 \\
\textbf{Full pipeline (proposed)} & \textbf{0.8367} & {0.9046} & \textbf{0.8629} \\
\bottomrule
\end{tabular}
\end{table}

The no-context-scoring variant achieves the highest recall (0.9093, bolded), but at a substantial precision cost (0.6495 vs.\ 0.8367), confirming that context scoring trades marginal recall for a large precision gain. The \textit{deterministic only} variant produces the lowest F1 (0.4057) due to coincidental numeric matches and the inability to recover multi-dimensional features. Removing \textit{engineering heuristics} preserves recall (0.8867) but collapses precision to 0.4423, because the system can no longer route dimension types to appropriate feature types. Disabling \textit{LLM escalation} yields good precision (0.7222) but poor recall (0.5417), confirming that VLM assistance is critical for resolving ambiguous cases. Removing \textit{context scoring} ($w_c = 0$) produces the best degraded F1 (0.7432) with high recall but reduced precision due to spurious dimensional agreements. The full pipeline achieves the best F1 (0.8629), 12.0 points above the nearest degraded variant.

\section{Discussion}

The proposed method achieves consistently high recall (90.46\%) across evaluated parts, indicating that the majority of ground truth correspondences are successfully recovered. The relatively low standard deviation in recall (11.65\%) suggests stable recovery performance across parts of varying complexity. In contrast, precision exhibits greater variance (mean 83.67\%, std 16.98\%), particularly in cases involving repeated or geometrically similar features. These scenarios highlight the inherent ambiguity of drawing-based intent specification and the practical limits of geometric disambiguation when contextual cues are insufficient.

Feature-level analysis further demonstrates the robustness of the approach. The high partial match rate (90.25\%) indicates stable behavior even when full disambiguation is not possible. In ambiguous cases the system favors retaining multiple plausible correspondences or deferring resolution rather than producing incorrect assignments, a policy essential for manufacturing-grade automation.

The ablation study (Table~\ref{tab:ablation}) confirms that each component contributes meaningfully, with the full pipeline achieving 12.0 F1 points above the nearest degraded variant. The resolution statistics (4.0\% deterministic, 84.1\% combined deterministic-VLM, 11.9\% LLM) corroborate this: VLM-assisted selection is needed for the majority of cases because contextual ambiguity prevents deterministic scoring alone from producing a single clear winner.

Several limitations should be acknowledged. First, the evaluation dataset is limited in size (20 parts) and is predominantly dimension-centric; GD\&T annotations are essentially absent, meaning the framework's GD\&T heuristics are not validated. Future evaluation on GD\&T-rich datasets is planned. Second, the high VLM involvement rate indicates that purely deterministic scoring rarely suffices, motivating investigation of richer spatial features to increase deterministic resolution. Third, mapping accuracy depends on upstream AFR and drawing entity extraction quality. Finally, VLM-based reasoning introduces computational overhead that may need optimization for high-throughput environments.

\section{Conclusion}

This paper presented a deterministic-first, context-aware framework for mapping 2D drawing entities to CAD features in support of manufacturing automation. By combining semantic enrichment, interpretable correspondence scoring, engineering-aware heuristics, and controlled escalation mechanisms, the proposed approach produced unified manufacturing specifications with explicit provenance, confidence estimates, and exception handling. Evaluation on 20 real CAD--drawing pairs achieved a mean F1 score of 86.29\%. Ablation experiments confirmed that each pipeline component contributes: removing engineering heuristics collapsed precision from 0.8367 to 0.4423, while removing LLM escalation reduced recall from 0.9046 to 0.5417, demonstrating the necessity of both domain knowledge and controlled AI-assisted escalation.

Future work will focus on scaling to larger assemblies, incorporating tolerance stack-up reasoning and inspection planning feedback, evaluating on GD\&T-rich datasets, and integrating the unified specification into end-to-end digital thread workflows. Overall, the emphasis on determinism, transparency, and manufacturing-grade reliability positions the proposed approach as a practical foundation for deploying AI-assisted design interpretation in industrial environments.

\section*{Acknowledgment}

This work is supported by the Singapore International Graduate Award (SINGA) (Awardee: Muhammad Tayyab Khan), funded by the Agency for Science, Technology and Research (A*STAR); an AcRF Tier 1 grant (RG70/23) from the Ministry of Education, Singapore, and Nanyang Technological University; and a Korea Planning \& Evaluation Institute of Industrial Technology (KEIT) grant funded by the Korean government (MOTIE) (No. RS-2024-00442448).

\bibliographystyle{IEEEtran}
\bibliography{references}

\end{document}